\documentclass[review]{elsarticle}
\usepackage{lineno,hyperref}
\pdfoutput=1
\modulolinenumbers[5]
\journal{Journal of \LaTeX\ Templates}

\begin{document}
\begin{frontmatter}
\title{Exactly solvable chain of interacting electrons with correlated hopping and pairing}
\tnotetext[mytitlenote]{Exactly solvable chain of interacting electrons with correlated hopping and pairing}
\author{Igor N.Karnaukhov}
\address{G.V. Kurdyumov Institute for Metal Physics, 36 Vernadsky Boulevard, 03142 Kiev, Ukraine}
\fntext[myfootnote]{karnaui@yahoo.com}



\begin{abstract}
A generalization of the Mattis-Nam model (J.Math.Phys., 13 (1972), 1185), which takes into account a correlated hopping and pairing of electrons, is proposed, its exact solution is obtained. In the framework of the model the stability of the zero energy Majorana fermions localized at the boundaries is studied in the chain in which electrons interact through both the on-site Hubbard interaction and the correlated hopping and pairing. The ground-state phase diagram of the model is calculated, the region of existence of topological states is determined. It is shown that low-energy excitations destroy bonds between electrons in the chain, leading to an insulator state.
\end{abstract}

\begin{keyword}
\texttt  Hubbard interaction \sep topological state \sep edge modes
\end{keyword}
\end{frontmatter}
\linenumbers

\section{Introduction}
Behavior of electrons interacting via short-range interaction is described in the framework of the Hubbard model, exact solution of which has been obtained in one dimension in \cite{LW} (see also \cite{KOR}). The Hubbard model with correlated hopping on a chain has been proposed and solved exactly in \cite{0,1,2,3}. Mattis and Nam (MN) proposed modification of the Hubbard model for interacting electrons forming pairs, and solved it exactly in special point \cite{MN} (better known as the Kitaev point \cite{K1}). In contrast to traditional Hubbard model \cite{LW}, the MN model describes topological states of interacting electrons \cite{K1,IK1,IK2}, quantum topological phase transition between topological trivial and nontrivial phases. Exact solvable (1+1)D models allow one to study topological states taking into account the interaction between particles, the presence of interaction can coordinately change the states in the system. A striking example of this is soliton excitations. In this context, it is interesting to discuss a new model, which is a modification \cite{1,MN}, the exact solution of which takes place for arbitrary on-site interaction, correlated hopping and pairing.

The realization of topological states in real systems is determined by the stability of the topological phase in the presence of interaction and disorder \cite{D1,D2,D3,D4,D5}.
Topological models that take into account the interaction, even richer than non-interacting, different phases and phase transitions can be realized in such models. Phases with topological order are not realized in one-dimensional systems in the absence of any symmetry, since all states in this case are trivial states. If there is a certain symmetry, there are phases with different topological orders that are protected by this symmetry \cite{W}.
The influence of interaction on the phase state in topological insulators remains largely unexplored, and our knowledge is largely limited. The main difficulty in studying the behavior of interacting topological systems is due to the strong correlations. There are no quantitatively reliable analytical methods for solving such tasks.

The Coulomb interaction between electrons destroys the topological state of the system, it 'kills' the topological state, the task is how to approach the solution of the key problem. Low-dimensional quantum models can be solved exactly at certain points, corresponding to defined values of the parameters. Using the example of a chain of interacting electrons, MN determined the region of stability of topological states with an arbitrary value of the interaction between electrons \cite{MN}.

In the paper, we considered an extended modification of the MN model, taking into account also the correlated hopping and pairing of electrons.
Using the MN approach, we will show, that the model has exact solution for arbitrary values of on-site Hubbard interaction and correlated hopping and pairing. On the phase diagram, topological trivial and nontrivial phases are separated by the lines of quantum topological phase transitions.   We have shown, that the effect of correlated hopping and pairing and on-site Hubbard interaction on the behavior of the electron chain is different.
Spinless fermions move in a static $Z_2$ gauge field, which is uniform in the ground state, similar to the Kitaev model \cite{K2,L}. `Defect` in the Z-configuration breaks two bonds between electron and its nearest-neighbors, forming an isolated electron state on the site. Such type of excitations leads to transition to insulator state of the chain.

\section{The model}

The Hamiltonian of the model is the sum of two terms ${\cal H}={\cal H}_{MN}+{\cal H}_{ch}$, the first of which is determined in accordance with the MN model, the second takes into account the correlated hopping  and pairing terms within this model
\begin{eqnarray}
&&{\cal H}_{MN}=
-\sum_{j=1}^{N-1}\sum_{\sigma=\uparrow,\downarrow}(c^\dagger_{j,\sigma}-c_{j,\sigma})( c^\dagger_{j+1,\sigma}+c_{j+1,\sigma})+ \nonumber\\
&&U\sum_{j=1}^{N}(n_{j,\uparrow}-\frac{1}{2}) (n_{j.\downarrow}-\frac{1}{2}),
\label{eq:H0}
\end{eqnarray}
\begin{eqnarray}
&&{\cal H}_{ch}=
\frac{i t_1}{2}\sum_{j=1}^{N-1}\sum_{\sigma =\uparrow,\downarrow} [
-(c^\dagger_{j,\sigma}+c_{j,\sigma})(c^\dagger_{j+1,\sigma}+c_{j+1,\sigma})(n_{j,-\sigma}-\frac{1}{2})+\nonumber\\
&&
(c^\dagger_{j,\sigma}-c_{j,\sigma})(c^\dagger_{j+1,\sigma}-c_{j+1,\sigma})(n_{j+1,-\sigma}-\frac{1}{2})]+
\nonumber\\&&
\frac{t_2}{4}\sum_{j=1}^{N-1}\sum_{\sigma =\uparrow,\downarrow}
(c^\dagger_{j,\sigma}+c_{j,\sigma})(n_{j,-\sigma}-\frac{1}{2})(c^\dagger_{j+1,\sigma}-c_{j+1,\sigma})(n_{j+1,-\sigma}-\frac{1}{2}),
\label{eq:H1}
\end{eqnarray}
where $c^\dagger_{j,\sigma},c_{j,\sigma} (\sigma=\uparrow,\downarrow)$ are the fermion operators determined on a lattice site $\emph{j}$,  $U$ is the  value of the on-site Hubbard interaction, $n_{j,\sigma}=c^\dagger_{j,\sigma}c_{j,\sigma}$, $t_1$ and $t_2$ determine the correlated hopping and pairing terms.
The Hubbard operators $c^\dagger_{j,\sigma}n_{j,-\sigma}$, $c_{j,\sigma}n_{j,-\sigma}$  determine the correlated hopping and pairing of electrons \cite{1,2,3}.

Using the Jordan-Wigner transformation for fermions with different spins, we determine the Hamiltonian (\ref{eq:H0})-(\ref{eq:H1})  via  spin-1/2 operators $\textbf{S}_j$ and $\textbf{T}_j$ for the particles with spin up and down \cite{MN}
$${\cal H}_{MN}= -4\sum_{j=1}^{N-1}(S_{j}^x S_{j+1}^x +T_j^x T_{j+1}^x)+U\sum_{j=1}^{N}S_j^zT_j^z$$ and
$${\cal H}_{ch}=  4t_1\sum_{j=1}^{N-1}(T_j^z S_j^y S_{j+1}^x +S_j^z T_{j}^y T_{j+1}^x-
S_j^x S_{j+1}^y T_{j+1}^z -T_j^x T_{j+1}^y S_{j+1}^z)$$
$$-4t_2\sum_{j=1}^{N-1}(T_j^z S_{j}^y S_{j+1}^y T_{j+1}^z +S_j^z T_{j}^y T_{j+1}^y S_{j+1}^z).$$
The authors \cite{MN} have introduced a new set of spin-1/2 matrices $\textbf{J}_j$ and $\textbf{P}_j$
($\{S^x_j,S^y_j,S^z_j\} = \{J^x_j, 2J^y_jP^x_j, 2J^z_jP^x_j\}$ and $\{T^x_j,T^y_j,T^z_j\}= \{-2P^z_jJ^x_j, 2P^y_jJ^x_j, P^x_j\}$), that gives possibility redetermine the total Hamiltonian in the following form
\begin{eqnarray}
&&{\cal H}=- \sum_{j=1}^{N-1}[4 J^x_j J^x_{j+1}+ t_2 J^y_j  J^y_{j+1} - 2t_1 (J^y_j J^x_{j+1} -J^x_j J^y_{j+1})](4 P^z_j P^z_{j+1}+1)\nonumber\\&&
+\frac{U}{2} \sum_{j=1}^{N}J^{z}_{j}
\label{eq:H2}
\end{eqnarray}
The $P^z_j$ operators commute with the total Hamiltonian, they are the integrals of the motion. The ground state energy of the spin chain corresponds to (\ref{eq:H2}) with $P^z_j P^z_{j+1}=\frac{1}{4}$, the ground state is twice degenerated. The $P^z_j$ operators form a static $Z_2$ gauge field, uniform configuration of this field corresponds to the ground state \cite{K2,L}.
The Hamiltonian $\cal H$ describes the XY-Heisenberg spin-$\frac {1}{2}$ chain in a magnetic field
\begin{eqnarray}
{\cal H}= -2\sum_{j=1}^{N-1}[ 4J^x_j J^x_{j+1}+t_2 J^y_jJ^y_{J+1}-2 t_1 (J^y_j J^x_{j+1} -J^x_j J^y_{j+1})]+\frac{U}{2} \sum_{j=1}^{N}J^{z}_{j}.
\label{eq:HI}
\end{eqnarray}
Redetermine the Hamiltonian $\cal H$ in the operators of spinless fermions $a^\dagger_j$ and $a_j$
\begin{eqnarray}
&&{\cal H}= -2\sum_{j=1}^{N-1}\left[ \left(1+i\frac{t_1}{2}+\frac{t_2}{4}\right)a^\dagger_{j+1}a_{j}+
\left(1- \frac{t_2}{4}\right)a^\dagger_j a^\dagger_{j+1}+H.c.\right]
\nonumber\\&&
+\frac{U}{2} \sum_{j=1}^{N}\left(a^\dagger_j a_j - \frac{1}{2}\right).
\label{eq:HF}
\end{eqnarray}

The model Hamiltonian (\ref{eq:H0})-(\ref{eq:H1}) defines a family of three parametric exactly solvable models which is determined by the parameters  $\{t_1,t_2,U\}$, the proposed models reduce to the MN model at $t_1=t_2=0$ and the Kitaev chain with zero chemical potential at $t_1=t_2=U =0$.

\section{The ground-state phase diagram}

\begin{figure}[tp]
     \centering{\leavevmode}
\begin{minipage}[h]{.75\linewidth}
\center{
\includegraphics[width=\linewidth]{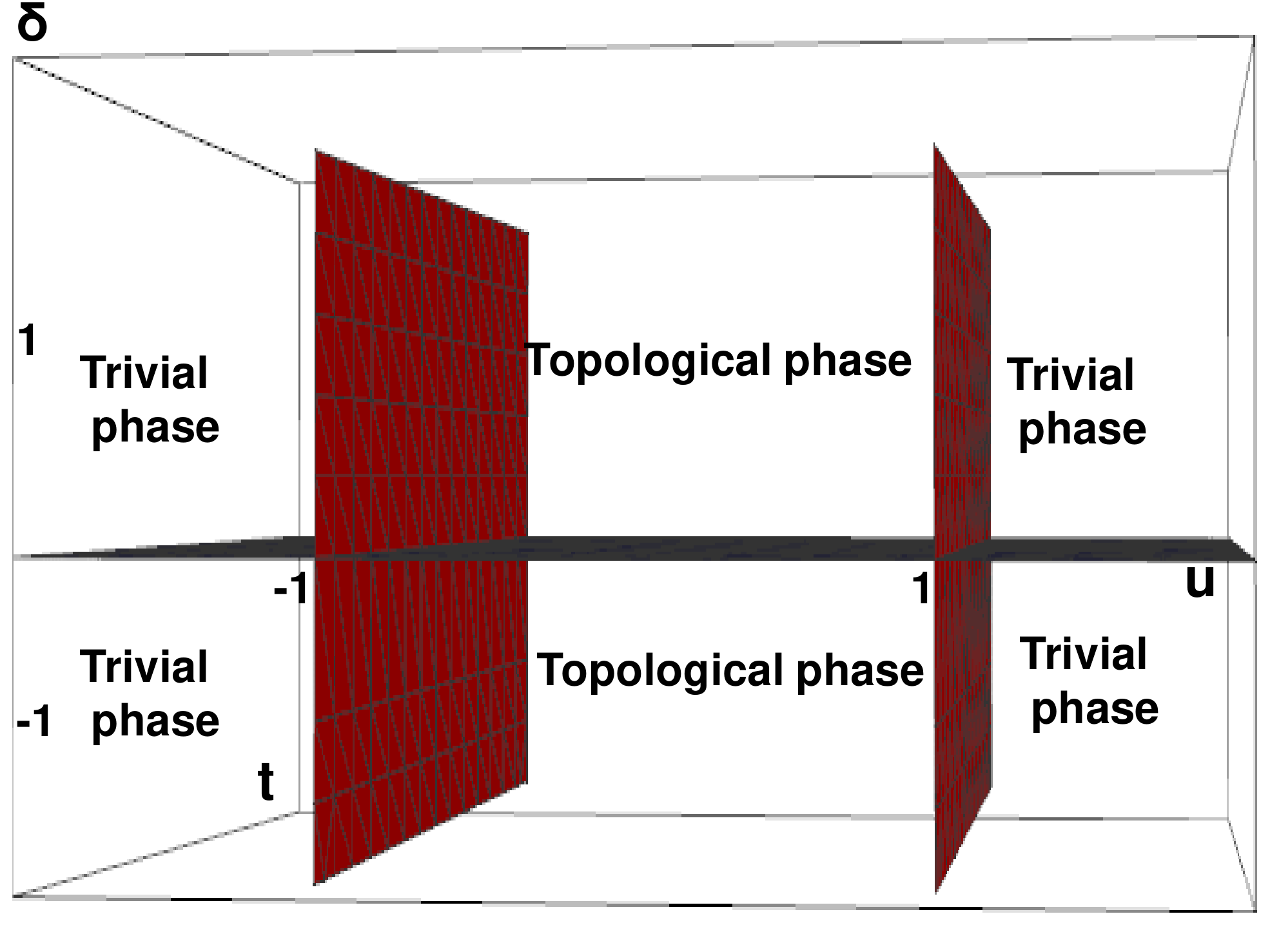}
                  }
    \end{minipage}
\caption{(Color online)
The ground-state phase diagram in the coordinates $u,t,\delta$:  at $|u|<1$ and $\delta \neq 0$ the wire is in topological phase with the  winding number $-\rm{sgn}\delta$, at $|u|>1$ in trivial phase, $\delta=0$ corresponds to metal phase, $u=\pm 1$ correspond to quantum phase transitions.
  }
\label{fig:1}
\end{figure}
The total Hamiltonian with the on-site Hubbard interaction and correlated hopping and pairing is mapped to a noninteracting model of spinless fermions, namely to the XY-Heisenberg spin-$\frac{1}{2}$ chain (\ref{eq:HF}). The spectrum of spinless fermions is symmetric with respect to zero energy, equal to
\begin{equation}
\epsilon (k) = \pm (2+t_2/2)\sqrt{\left(u-\cos k -t\sin k \right)^2+\delta^2 \sin ^2 k},
\label{eq:H3}
\end{equation}
where $k$ is the wave vector of fermion excitations along the chain, $u=\frac{U}{8+2t_2}$, $t=\frac{t_1}{2+t_2/2}$, $\delta=\frac{1-t_2/4}{1+t_2/4}$.

At the points of the topological phase transition  $u=\pm 1$ the excitation spectrum gap is equal to zero, the points of the phase transitions separate trivial and nontrivial topological phases.  Let us define the winding number $\nu$ via the angle $\theta(k)$ that determines the Bogoliubov transformation in superconductors
\begin{equation}
\nu=\frac{1}{2\pi}\int_{k \in BZ}\frac{\partial\theta(k)}{\partial k}dk =-\rm{sgn}\delta.
\label{eq:wn}
\end{equation}
where integration is carried out over the Brillouin zone, $\tan \theta(k)=\frac{\delta\sin k}{u-\cos k-t\sin k}$.

At $|u|>1$ the system is in trivial topological state, for  $|u|<1$ and $\delta\neq 0$ it is in topological state, which is determined by the  winding number $\pm 1$ (\ref{eq:wn}). In trivial state the winding number is equal to zero, $\delta=0$ corresponds to normal metal state, that is gapless for $|u| <1$. Note, that the phase diagram does not depend on the value of correlated hopping and pairing $t_1$ (see in Fig.1).
The topological state is characterized by the zero energy Majorana fermions localized at the boundaries \cite{K1}. The quantum Ising model  (\ref{eq:HI}) equivalent to model of free spinless fermions  (\ref{eq:HF}) \cite{a,b,c}.
The trivial topological phase corresponds to a paramagnetic phase polarized in the $\sigma^z$-direction. The two topological phases correspond to an anti-ferromagnet for $t_2/2-2>0$ in the $\sigma^y$-direction or in the $\sigma^x$-direction for $t_2/2-2<0$. These  anti-ferromagnetic phases are not topologically protected, while the ground state is two-fold degenerate.

In topological trivial phase zero energy Majorana fermions are absent. The on-site Hubbard interaction kills the zero energy Majorana fermions at $|U|>8+2t_2$, it limits  the ambitions of topological phase in the chain with strong interaction between electrons. The correlated hopping and pairing considered in the model Hamiltonian expands the region of existence of the topological phase when it has the same sign with the single-particle hopping integral, and decreases it when they have different signs. Near the point $t_2=-4$ the topological state is not stable with respect to small fluctuations of the on-site Hubbard interaction, small fluctuations of $U$ open the gap in the fermion spectrum, stabilizing the topological trivial state in the chain.

\subsection{Zero-energy edge states of Majorana fermions}
\begin{figure}[tp]
     \centering{\leavevmode}
     \begin{minipage}[h]{.75\linewidth}
\center{
\includegraphics[width=\linewidth]{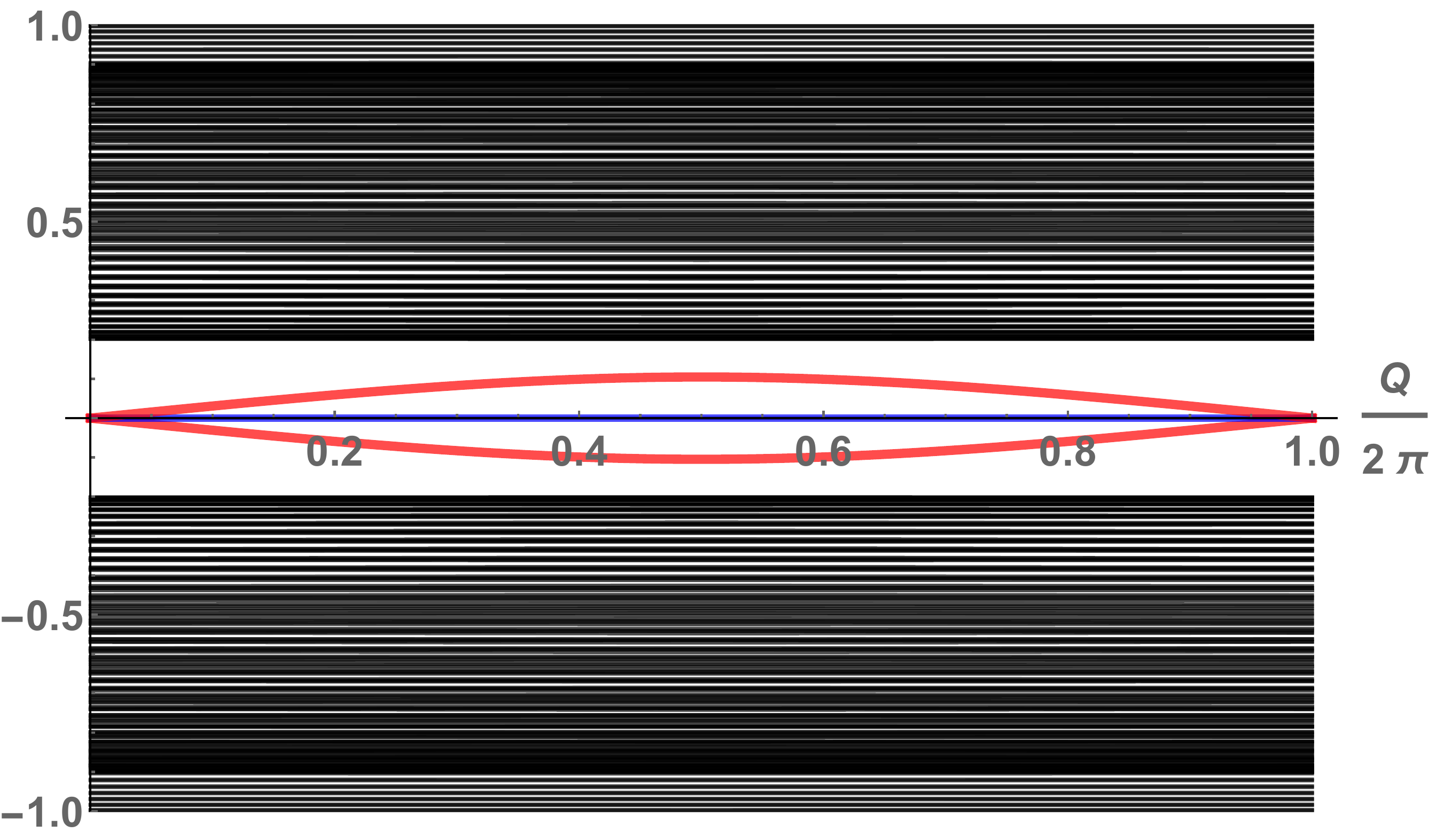} a)\\
                  }
    \end{minipage}
     \centering{\leavevmode}
\begin{minipage}[h]{.75\linewidth}
\center{
\includegraphics[width=\linewidth]{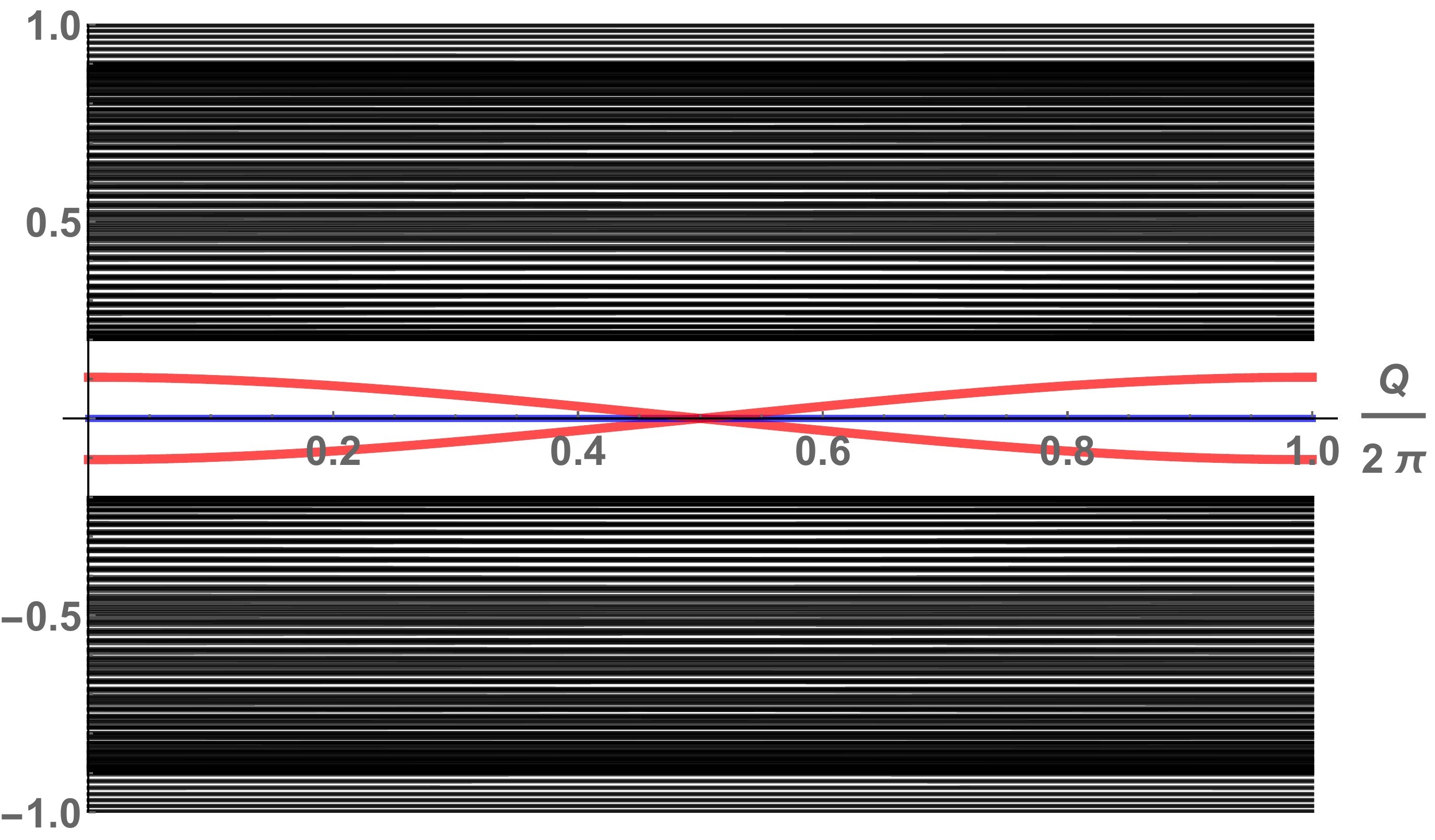} b)\\
                  }
    \end{minipage}
\caption{(Color online)
Low energy spectrum of the chain $\frac{\epsilon}{2+t_2}$ as function of $Q$  calculated at $u=t=\tau=0.2$ a) $\delta_L=\delta_R=0.2$ b) $\delta_L=-\delta_R=0.2$. The cases a) and b) correspond to $\gamma_R-\chi_L$ ($\gamma_L-\chi_R$) and  $\gamma_R-\gamma_L$ ( $\chi_L-\chi_R$) configurations of Majorana fermions on the junction, respectively.  Majorana fermion states localized at the junction are marked in red and at free boundaries in blue, the spectrum is calculated in the chains length $L=200$ with open boundary conditions.}
\label{fig:2}
\end{figure}

Two zero-energy Majorana fermions localized at the boundaries have a different structure. Below we consider the possibility of testing the types of the Majorana fermions, that are realized at the boundary of a superconducting wire. Let us consider two wires (denotes as L and R) connected by the term ${\cal H}_{tun}$, which contains the tunneling amplitude $\tau$ and takes into account the applied flux $Q$
\begin{equation}
 {\cal H}_{tun}= 2 \tau \exp (iQ/2) c^\dagger_L c_R+ H.c.,
\label{eq:H4}
\end{equation}
where the Fermi operators $c_L$, $c_R$ determine the tunneling of fermions between two superconducting wires.

The fermion operators $c_L$, $c_R$ can be represented as a sum of two Majorana fermions $\gamma_{L,R}$, $\chi_{L,R}$: $\gamma_{L,R}=c_{L,R}+c^\dagger_{L,R}$ and  $\chi_{L,R}=i(c^\dagger_{L,R}-c_{L,R})$. The Hamiltonian (\ref{eq:H4}) can be straightforwardly rewritten in terms of these operators \cite{4}:
\begin{equation}
{\cal H}_{tun}= \tau \sin (Q/2 )(\gamma_L\gamma_R +\chi_L \chi_R)+ i\tau \cos(Q/2)(\gamma_L\chi_R - \chi_L \gamma_R).
\label{eq:H5}
\end{equation}

According to (\ref{eq:H5}) the tunneling Majorana fermions between wires depends on the  $\gamma,\chi$-zero energy Majorana fermions localized at the junction, as a result, a $4\pi$-persistent current through the junction \cite{K1,D4,D5,4} is shifted on $\pi$ for $\gamma_L - \chi_R, \chi_L - \gamma_R$ or $\gamma_L - \gamma_R, \chi_L - \chi_R$ configurations of Majorana fermions localized at the junction.
Numerical calculations of the low-energy excitation spectrum of two superconducting wires described by the Hamiltonian (\ref {eq:HF}) with $ \delta_L = \delta_R $ and $ \delta_L = - \delta_R $, connected by the Hamiltonian (\ref {eq:H5}) are shown in Fig.2 (for illustration above). The bulk spectrum of superconductors is the same, the behavior of the edge states is different, since the winding numbers have different signs. The persistent current through the junction is determined by the Majorana edge states localized at the junction (marked in red in Fig.2)).

\subsection{Insulator state as a result of excitations in the chain}

At half filling the Fermi energy is equal to zero, electrons are paired in the pairs which form condensate. Consider excitations in topological state, as a example the case $U=2$, $t_1=0$, $t_2=1$ is shown in Fig.3. In this case the energy of isolated electron, equal to $\frac{U}{2}$, lies into the gap. Hight energy subband corresponds to one-particle excitations of electrons. As we note in previous section the ground state of the chain corresponds to uniform configurations of $P^z_j$ operators, which are integrals of motion $P^z_j=\frac{1}{2}$ or $P^z_j=-\frac{1}{2}$. We show, that in the chain with periodic boundary conditions, one '$P^z_l$-defect' at $l$-site ($P^z_l\neq P^z_j$) leads to an insulator state of the chain. According to (\ref{eq:H2}) one '$P^z_l$-defect' forms a fermion, isolated on $l$-site, as a result, we obtain the chain  with open boundary conditions. One electron pair decays into one electron with energy $\frac{U}{2}$, localized at $l$-site, and zero energy Majorana fermions localized at the boundaries $l-1$ and $l+1$. This state, which has a minimum excitation energy $\frac {U} {2} < \frac{\Delta}{2} $, corresponds to the breaking of bonds between electrons in a chain (between $l-1$ and $l+1$ sites), triggers insulator state of the chain.

\begin{figure}[tp]
     \centering{\leavevmode}
\begin{minipage}[h]{.75\linewidth}
\center{
\includegraphics[width=\linewidth]{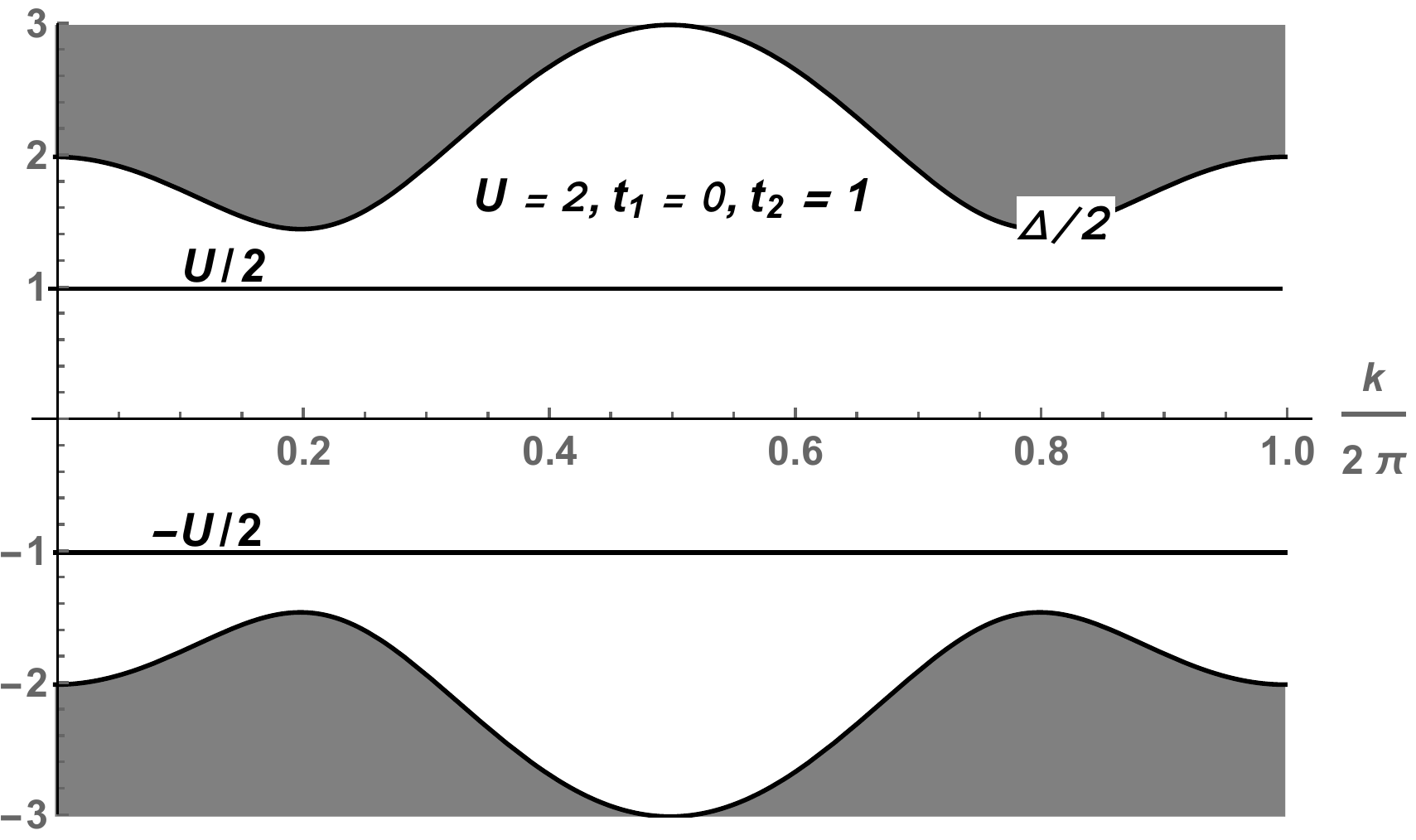}
                  }
    \end{minipage}
     \centering{\leavevmode}
\caption{The spectrum of chain as a function of the wave vector $k$  in the topological state for $U=2$, $t_1=0$, $t_2=1$ with one '$P^z_l$-defect' at $l$-site ($P^z_l\neq P^z_j$).
}
\label{fig:3}
\end{figure}

\section{Conclusion}

We have considered the exact solution of a three-parameter family of 1D models of electrons interacting via the on-site Hubbard repulsion and correlated hopping and pairing. The model Hamiltonian reduces to non-interacting spinless fermions hopping and pairing in the background of static $Z_2$ gauge field configurations, the ground state corresponds to uniform configurations. It is shown that the excitations corresponding to the defect in the static field configuration leads to breaking of the bonds between the electrons, induces the dielectric state of the chain. The ground-state phase diagram includes topological trivial and nontrivial phases. Criteria of realization topological state, which characterized by zero energy Majorana fermions localized at the boundaries are calculated for arbitrary values of the on-site Hubbard interaction and integrals of correlated hopping and pairing.

This research was partially supported by the budget program 6541230 "Support for the development of priority areas of research".

\section*{References}
\bibliography{mybibfile}

\end{document}